\documentclass[10pt,a4paper]{article}

\usepackage{graphicx}
\usepackage{epsfig}
\usepackage{amsmath}

\newcommand{\be}{\begin{equation}}
\newcommand{\ee}{\end{equation}}
\newcommand{\ba}{\begin{equation} \begin{aligned}}
\newcommand{\ea}{\end{aligned} \end{equation}}
\newcommand{\num}[1]{\left[ #1 \right]}

\newcommand{\SIR}{{\it SIR}}

\newcommand{\SIS}{{\it SIS}}

\newfont{\mtfont}{newmotifsymbols}
\newcommand{\mtI}{\mbox{\mtfont B}}
\newcommand{\mtII}{\mbox{\mtfont C}}
\newcommand{\mtIII}{\mbox{\mtfont D}}
\newcommand{\mtIV}{\mbox{\mtfont E}}
\newcommand{\mtV}{\mbox{\mtfont F}}
\newcommand{\mtVI}{\mbox{\mtfont G}}
\newcommand{\mtPair}{\mbox{\mtfont H}}
\newcommand{\mtTriangle}{\mbox{\mtfont I}}
\newcommand{\mtTriple}{\mbox{\mtfont J}}
\newcommand{\mtNode}{\mbox{\mtfont K}}

\makeatletter
\newdimen\myht
\def\mybrackskip{\hskip.3em}

\def\mybrack#1#2{%
{\vrule height #1 width \myht}%
{\leaders\hrule height #2\hfill}%
{\vrule height #1 width \myht}%
}

\def\mydotbrack#1#2{%
{\vrule height #1 width \myht}%
\dotfill%
{\vrule height #1 width \myht}%
}

\def\myfourmotif#1#2#3#4#5#6#7#8{%
\setbox0=\hbox{$#1$}%
\setbox1=\hbox{$#2$}%
\setbox2=\hbox{$#3$}%
\setbox3=\hbox{$#4$}%
\setbox4=\hbox{--}%
\def\mybracki{%
\ifnum#5=2 \mydotbrack{0pt}{\myht} \fi \ifnum#5=1
\mybrack{0pt}{\myht} \fi }
\def\mybrackii{%
\ifnum#5=2 \mybrack{2pt}{0pt} \fi \ifnum#5=1 \mybrack{2pt}{0pt} \fi
\ifnum#5=0 \mybrack{0pt}{0pt} \fi }
\def\mybrackiii{%
\ifnum#7=2 \mydotbrack{2pt}{\myht} \fi \ifnum#7=1
\mybrack{2pt}{\myht} \fi }
\def\mybrackiv{%
\ifnum#8=2 \mydotbrack{4pt}{\myht} \fi \ifnum#8=1
\mybrack{4pt}{\myht} \fi }
\def\myraise{%
\ifnum#5>0 9.5pt \else 7.5pt \fi }
\def\mylink{%
\ifnum#6=2 $\cdots$%
\fi \ifnum#6=1 \copy4 \fi \ifnum#6=0 \hskip\wd4 \fi }
\mathop{\raise \myraise\vtop{\m@th\ialign{##&##&##&##&##&##&##\crcr%
\crcr& & \hskip.5\wd1\mybracki \span \span \span &%
\crcr\noalign{\kern 0pt \nointerlineskip}%
\crcr& & \hskip.5\wd1\mybrackii \span \span \span &%
\crcr\noalign{\kern 0pt \nointerlineskip}%
$\lbrack$\mybrackskip%
\hfil&\copy0 --&\copy1 &--&\copy2 &\mylink\copy3 &\hfil%
\mybrackskip$\rbrack$%
\crcr\noalign{\kern 0pt\nointerlineskip}%
&\hskip.5\wd0\mybrackiii \span \span \span & &%
\crcr\noalign{\kern -2pt\nointerlineskip}%
&\hskip.5\wd0\mybrackiv \span \span \span \span &%
\crcr%
}}}}

\def\mythreemotif#1#2#3#4{%
\setbox0=\hbox{$#1$}%
\setbox1=\hbox{$#2$}%
\setbox2=\hbox{$#3$}%
\mathop{\lower 0pt\vtop{\m@th\ialign{##&##&##&##&##\crcr%
$\lbrack$\mybrackskip%
\hfil&\copy0 --&\copy1 &--\copy2 &\hfil%
\mybrackskip$\rbrack$%
\crcr\noalign{\kern -1pt\nointerlineskip}%
\ifnum#4=2
&\hskip.5\wd0\mydotbrack{2pt}{\myht}\hskip.5\wd2\span\span&\crcr \fi
\ifnum#4=1
&\hskip.5\wd0\mybrack{2pt}{\myht}\hskip.5\wd2\span\span&\crcr \fi
}}}}

\def\mypair#1#2{%
\hbox{$\lbrack$\mybrackskip$#1$--$#2$\mybrackskip$\rbrack$}}

\def\mynode#1{\hbox{$\lbrack$\mybrackskip$#1$\mybrackskip$\rbrack$}}
\makeatother

\begin{document}

\title{Generalised network clustering and its dynamical implications}

\author{Thomas House}


\date{}

\maketitle

\begin{abstract}

A parameterisation of generalised network clustering, in the form of four-motif
prevalences, is presented. This involves three real parameters that are
conditional on one- two- and three-motif prevalences. Interpretations of these
real parameters are presented that motivate a set of rewiring schemes to create
appropriately clustered networks. Finally, the dynamical implications of higher
order structure, as parameterised, for a contact process are considered.

\end{abstract}

\section{Introduction}

Networks have become one of the indispensible tools for the study of complex
systems with many interacting components, as demonstrated by their ubiquity in
the Proceedings of the recent European Conference on Complex Systems with which
this journal issue is concerned. In particular, the combination of high
clustering amongst nodes and short average path length, commonly known as the
small world phenomenon~\cite{Watts:1998}, has been observed not only in
social networks~\cite{wassermanFaust94}, but also in technological, metabolic
and citation networks~\cite{Newman:2003p1751,Strogatz:2001p1750,Watts:2003}.

Small connected sub-graphs of complex networks, known as \emph{motifs}, have
also been observed to have significantly different prevalences from those
expected in a random case, leading to scientific
insight~\cite{Milo:2002p1689,Milo:2004p8}. This paper is concerned with an
alternative approach to motif prevalence that conditions on standard,
triangle-level clustering, as a guide to intuition for other applications of
the concept of motifs. In particular, new wirings are presented that modify
clustering without changing node degree, along the lines
of~\cite{Bansal2009,Kiss:2008p1881,House:pcbi,Green:2009p1883}.

Networks, and population structure in general, have also become central to
modern infectious disease epidemiology~\cite{KRbook}. The impact of motif
structure for \SIS{} epidemics was considered in~\cite{House:2009p1790}, and we
combine the dynamical system developed in that work with the new
parameterisation to gain insights into the impact of higher-order clustering on
transmission / contact process dynamics.

\section{Characterisation of motif structure}

We start by considering the relatively simple structure of one- two- and
three-motif prevalences. At orders one and two, there are only the number of
nodes in a network and the number of links to consider. For simplicity, we
consider networks with a single giant component of $N$ nodes in which each
individual has exactly $n$ links connecting it to the rest of the network.
This assumption is not essential to the general thrust of analysis presented,
but does simplify an already complex set of manipulations.  In our notation, we
use a diagramatic representation of a node and linked nodes enclosed in square
brackets to denote prevalence of that motif in the network.  This means that at
order one and two:
\be
\num{\mtNode} = N \; , \quad
\num{\mtPair} = nN \; .
\ee
So the motif structure at this level is given equivalently either by the raw
motif prevalences $\num{\mtNode}, \num{\mtPair} $, or by the real numbers $N,
n$. The benefit of the latter approach is that $n$ tells us something about the
number of links \emph{per node}---i.e.\ two-motif structure conditional on
one-motif structure.

Less trivially, there are two connected three-motifs: triangles and unclosed
triples. Since every triple must be either closed or unclosed, the prevalences
of three-motifs, notated using square brackets and diagrams as for other
motifs, obey the identity
\be
\num{\mtTriple} + \num{\mtTriangle} = Nn(n-1) \; .
\label{triident}
\ee
This means that a real parameter  $\phi\in [ 0, 1]$ can be introduced to
partition this identity as below:
\be
\num{\mtTriple} =  N n (n-1)(1-\phi ) \; , \quad
\num{\mtTriangle} = N n (n-1)\phi \; .
\label{tripartition}
\ee
In network analysis, $\phi$ (the ratio of triangles to all triples, closed and
unclosed) is often called the \textit{clustering coefficient}. In the same way
that $n$ conditions on network size, $\phi$ conditions on network size and
number of links to measure transitivity of the network in a different manner
from raw counts of triangles.

We now attempt a similar parameterisation at order four.  There are six
connected graphs of size four, which can be represented pictorally using the
following set of symbols:
$$
 \left\{ \mtI , \mtII , \mtIII , \mtIV , \mtV , \mtVI \right\} \text{ .}
$$
A set of identities analagous to~\eqref{triident} was introduced
in~\cite{House:2009p1790},
\ba
\num{\mtIII} + 2 \num{\mtV} + \num{\mtVI}
& = (n-2)\num{\mtTriangle} \; , \\
\num{\mtI} + 2\num{\mtIII} + \num{\mtV} 
& = (n-2)\num{\mtTriple} \; , \\
\num{\mtII} + \num{\mtIII} + \num{\mtIV} + \num{\mtV} 
& = (n-1)\num{\mtTriple} \; .
\label{idents}
\ea
Each of these identies is derived by starting with the three-motif appearing on
the right-hand side of the identity (either a triangle or unclosed triple) and
then joining a fourth node to one of the original three; the left-hand side of
each identity can then be seen as an enumeration of the possible additional
links between the new node and the two other nodes within the original
three-motif.  We now propose the main innovation of this work, a partition of
these identities in terms of three real parameters, $\psi, \zeta, \xi$.

We start this process by writing down the four-motif prevalences that would be
expected if transitive closure of any given triple is a random event of
constant probability $\phi$. In the case where no triangles at all are present
in the network, the only four-motif clustering structure possible is the
closure of four-lines into squares, and the appropriate motifs obey
$\num{\mtII} + \num{\mtIV} = Nn(n-1)^2$. This motivates the introduction of a
square-level partition of these two motifs, $\psi$, analagous to $\phi$
in~\eqref{tripartition}, but not equivalent in the case where some triangles
are present in the network.

\begin{figure}[H]
\begin{center}
\scalebox{0.9}{\resizebox{\textwidth}{!}{ \includegraphics{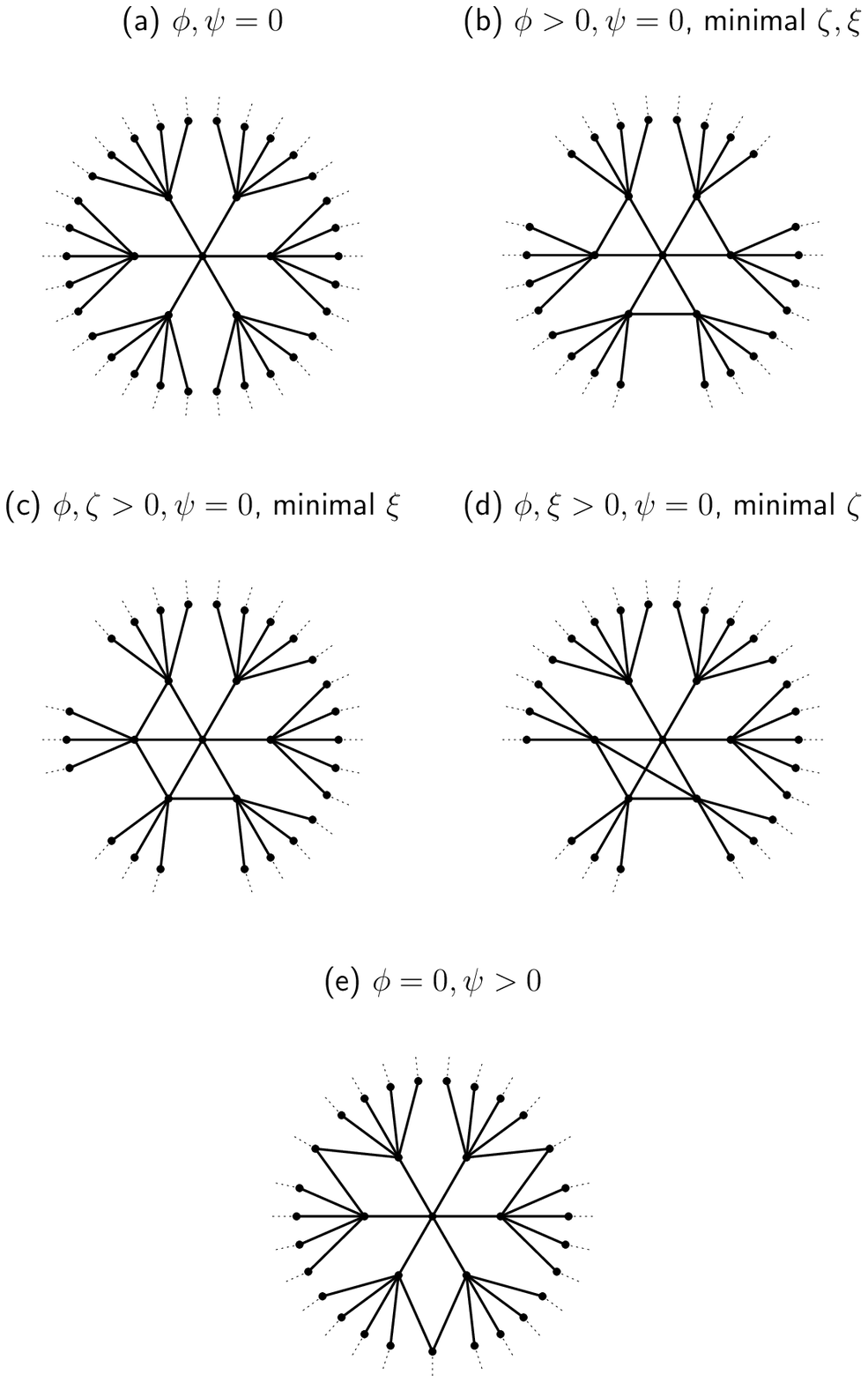} }}
\end{center}
\caption{Interpretation of the clustering parameters $\phi, \psi, \zeta, \xi$
for a typical neighbourhood in a network with $n=6$}
\label{fig:interpret}

\newpage

\end{figure}

Finally, we introduce parameters $\zeta$ and $\xi$ additively to the
prevalences of the motifs $\mtV$ and $\mtVI$ respectively, and then use the
identities~\eqref{idents} to carry through the consequences of this addition
to other motif prevalences, yielding the form
\ba
\num{\mtI} & = 
  N n (n-1)(n-2)\left((1-\phi)^3 + 3 \zeta \right)\; , \\
\num{\mtII} & = N n (n-1) \left( (n-1)(1-\phi) 
   - (n-2)\left(\phi(1 - \phi) - \zeta - \frac13 \xi \right) \right) (1-\psi)
  \; , \\
\num{\mtIII} & = 
  N n (n-1)(n-2) \left(\phi (1-\phi)^2 - 2 \zeta + \frac13 \xi \right)\; , \\
\num{\mtIV} & = N n (n-1) \left( (n-1)(1-\phi) 
   - (n-2)\left(\phi(1 - \phi) - \zeta - \frac13 \xi \right) \right) \psi
  \; , \\
\num{\mtV} & = 
  N n (n-1)(n-2) \left(\phi^2 (1-\phi) + \zeta - \frac23 \xi \right)\; , \\
\num{\mtVI} & = 
  N n (n-1)(n-2) \left(\phi^3 + \xi \right)\; .
\label{newparams}
\ea
Requiring that no motif prevalence be negative, the new parameters sit in the
following ranges, provided each of the others is zero:
\ba
\psi & \in [ 0 , 1] \; , \\
\zeta & \in \left[\max\left(-\frac13(1-\phi)^3, -\phi^2(1-\phi)\right) , 
  \frac12\phi(1-\phi)^2\right] \; , \\
\xi & \in \left[\max\left(-3\phi(1-\phi)^2, -\phi^3 \right) , 
  \frac32\phi^2(1-\phi)\right] \; .
\ea
A neighbourhood-based interpretation of these new parameters for certain
limiting cases is considered in Figure~\ref{fig:interpret}. This figure shows a
typical neighbourhood around an individual in a network with $n=6$, and
clustering parameter values varied.  Plot (a) shows a completely unclustered
graph---essentially a Cayley tree of degree 6. In (b), triangle-level
clustering $\phi$ has been introduced, but in such a way that the triangles do
not form highly connected fourth-order structures. (c) shows how the `envelope'
shape $\mtV$ is more prevalent than would be expected on the basis of the
three-motif structure: this means that $\zeta$ is positive, while for
under-represented envelopes $\zeta$ would be negative. (d) shows that the
`four-clique' $\mtVI$ is, in the same way, over-represented, implying positive
$\xi$, while its under-representation would imply negative $\xi$. Finally, (e)
shows that $\psi$ represents the ratio of squares to all four-lines (closed and
unclosed), and involves connections being made further away from the central
node than other clustering parameters. In this plot, as with (b), the squares
are shown maximally uncorrelated---obviously, at still higher orders of
clustering, correlations between squares may be parameterised as for triangles.

Outside of these limiting cases, however, the interpretation of the new
clustering parameters is more subtle, since the consistency
conditions~\eqref{idents} are much more structured than~\eqref{triident}.  In
this sense, the parameterisation of four-motif structure is not a
straightforward extension of the methodology used at the three-motif level.

\section{Rewiring schemes}

Rewiring schemes that preserve the number of links attached to a node can play
an important role in understanding, creating and manipulating networks. We now
present a set of rewiring schemes that modify the clustering parameters we have
introduced, two from existing work (together with applications) and three that
are, to our knowledge, novel.  These rewirings are an aid to intuition and also
demonstrate that explicit networks of the kind considered here can be generated
given sufficient computational resources.  Nevertheless, their na\"{i}ve
implementation is highly computationally intensive, and does not scale well
with network size, meaning that technical innovation beyond the scope of this
paper is necessary to produce simulations equivalent to the results obtained
below using moment closure.

\subsection{Randomiser}
This rewiring was used recently in epidemiological
applications~\cite{House:pcbi,Green:2009p1883} to remove all forms of
clustering without changing degree distribution.
\begin{center}
\begin{picture}(300,80)(0,0)
\put (20,20){\makebox(0,0){$i$}}
\put (60,20){\makebox(0,0){$k$}}
\put (20,60){\makebox(0,0){$j$}}
\put (60,60){\makebox(0,0){$l$}}
\put (20,25){\line(0,1){30}}
\put (60,25){\line(0,1){30}}
\put (100,40){\makebox(0,0){$\longrightarrow$}}
\put (140,20){\makebox(0,0){$i$}}
\put (180,20){\makebox(0,0){$k$}}
\put (140,60){\makebox(0,0){$j$}}
\put (180,60){\makebox(0,0){$l$}}
\put (145,20){\line(1,0){30}}
\put (145,60){\line(1,0){30}}
\end{picture}
\end{center}

\subsection{`Big V'}
This rewiring was considered recently
in~\cite{Bansal2009,Kiss:2008p1881,House:pcbi} to increase $\phi$ without
changing the degree distribution. 
\begin{center}
\begin{picture}(300,80)(0,0)
\put (60,20){\makebox(0,0){$O$}}
\put (40,40){\makebox(0,0){$a$}}
\put (20,60){\makebox(0,0){$A$}}
\put (80,40){\makebox(0,0){$b$}}
\put (100,60){\makebox(0,0){$B$}}
\put (25,55){\line(1,-1){12}}
\put (43,37){\line(1,-1){13}}
\put (77,37){\line(-1,-1){13}}
\put (95,55){\line(-1,-1){12}}
\put (120,40){\makebox(0,0){$\longrightarrow$}}
\put (180,20){\makebox(0,0){$O$}}
\put (160,40){\makebox(0,0){$a$}}
\put (140,60){\makebox(0,0){$A$}}
\put (200,40){\makebox(0,0){$b$}}
\put (220,60){\makebox(0,0){$B$}}
\put (145,60){\line(1,0){70}}
\put (163,37){\line(1,-1){13}}
\put (197,37){\line(-1,-1){13}}
\put (165,40){\line(1,0){30}}
\end{picture}
\end{center}

\subsection{`Big U'}
This novel rewiring increases $\psi$ 
\begin{center}
\begin{picture}(300,80)(0,0)
\put (40,20){\makebox(0,0){$O$}}
\put (80,20){\makebox(0,0){$\Theta$}}
\put (40,40){\makebox(0,0){$a$}}
\put (40,60){\makebox(0,0){$A$}}
\put (80,40){\makebox(0,0){$b$}}
\put (80,60){\makebox(0,0){$B$}}
\put (45,20){\line(1,0){30}}
\put (40,26){\line(0,1){10}}
\put (80,26){\line(0,1){8}}
\put (40,45){\line(0,1){10}}
\put (80,45){\line(0,1){10}}
\put (120,40){\makebox(0,0){$\longrightarrow$}}
\put (160,20){\makebox(0,0){$O$}}
\put (200,20){\makebox(0,0){$\Theta$}}
\put (160,40){\makebox(0,0){$a$}}
\put (160,60){\makebox(0,0){$A$}}
\put (200,40){\makebox(0,0){$b$}}
\put (200,60){\makebox(0,0){$B$}}
\put (160,26){\line(0,1){10}}
\put (200,26){\line(0,1){8}}
\put (165,60){\line(1,0){30}}
\put (165,40){\line(1,0){30}}
\put (165,20){\line(1,0){30}}
\end{picture}
\end{center}

\subsection{`YV'}
This novel rewiring increases $\zeta$ and $\phi$.
\begin{center}
\begin{picture}(300,80)(0,0)
\put (20,20){\makebox(0,0){$O$}}
\put (20,40){\makebox(0,0){$a$}}
\put (20,60){\makebox(0,0){$A$}}
\put (60,20){\makebox(0,0){$c$}}
\put (60,40){\makebox(0,0){$b$}}
\put (60,60){\makebox(0,0){$B$}}
\put (100,20){\makebox(0,0){$C$}}
\put (100,40){\makebox(0,0){$B'$}}
\put (25,23){\line(2,1){30}}
\put (25,20){\line(1,0){30}}
\put (65,20){\line(1,0){30}}
\put (65,40){\line(1,0){30}}
\put (20,26){\line(0,1){10}}
\put (20,45){\line(0,1){10}}
\put (60,45){\line(0,1){10}}
\put (140,40){\makebox(0,0){$\longrightarrow$}}
\put (180,20){\makebox(0,0){$O$}}
\put (180,40){\makebox(0,0){$a$}}
\put (180,60){\makebox(0,0){$A$}}
\put (220,20){\makebox(0,0){$c$}}
\put (220,40){\makebox(0,0){$b$}}
\put (220,60){\makebox(0,0){$B$}}
\put (260,20){\makebox(0,0){$C$}}
\put (260,40){\makebox(0,0){$B'$}}
\put (185,23){\line(2,1){30}}
\put (180,26){\line(0,1){10}}
\put (220,25){\line(0,1){10}}
\put (260,25){\line(0,1){9}}
\put (185,20){\line(1,0){30}}
\put (185,40){\line(1,0){30}}
\put (185,60){\line(1,0){30}}
\end{picture}
\end{center}

\subsection{`YYY'}
This novel rewiring increases $\xi$ and $\phi$.

\hspace{-.7cm}
\begin{picture}(365,120)(0,0)
\put (0,40){\makebox(0,0){$A$}}
\put (40,20){\makebox(0,0){$A'$}}
\put (40,40){\makebox(0,0){$a$}}
\put (40,100){\makebox(0,0){$B$}}
\put (80,60){\makebox(0,0){$O$}}
\put (80,80){\makebox(0,0){$b$}}
\put (120,20){\makebox(0,0){$C$}}
\put (120,40){\makebox(0,0){$c$}}
\put (120,100){\makebox(0,0){$B'$}}
\put (160,40){\makebox(0,0){$C'$}}
\put (45,98){\line(2,-1){30}}
\put (45,43){\line(2,1){29}}
\put (85,57){\line(2,-1){30}}
\put (84,83){\line(2,1){29}}
\put (5,40){\line(1,0){30}}
\put (125,40){\line(1,0){28}}
\put (40,26){\line(0,1){10}}
\put (80,65){\line(0,1){10}}
\put (120,26){\line(0,1){10}}
\put (180,70){\makebox(0,0){$\longrightarrow$}}
\put (200,40){\makebox(0,0){$A$}}
\put (240,20){\makebox(0,0){$A'$}}
\put (240,40){\makebox(0,0){$a$}}
\put (240,100){\makebox(0,0){$B$}}
\put (280,60){\makebox(0,0){$O$}}
\put (280,80){\makebox(0,0){$b$}}
\put (320,20){\makebox(0,0){$C$}}
\put (320,40){\makebox(0,0){$c$}}
\put (320,100){\makebox(0,0){$B'$}}
\put (360,40){\makebox(0,0){$C'$}}
\put (245,100){\line(1,0){68}}
\put (243,45){\line(1,1){33}}
\put (245,43){\line(2,1){29}}
\put (285,57){\line(2,-1){30}}
\put (284,78){\line(1,-1){33}}
\put (205,38){\line(2,-1){30}}
\put (245,40){\line(1,0){69}}
\put (280,65){\line(0,1){10}}
\put (325,23){\line(2,1){28}}
\end{picture}

\section{Contact-process dynamics}

We now present the model of~\cite{House:2009p1790}, used to investigate the
impact of higher order-clustering on what epidemiologists call \SIS{} dynamics.
In these dynamics, often called a~\textit{contact process}, individuals are
either susceptible ($S$) or infectious ($I$) with letters $A,B,C\ldots$
representing either of these states. Transmission of infection happens between
infectious individuals $I$ and susceptible individuals $S$ linked on the
network at a rate $\tau$, while infectious individuals recover and become
susceptible, since recovery is assumed not to offer lasting immunity, at a rate
$g$. We use square brackets to denote the prevalence of certain structures in
the network.

\subsection{Exact dynamical equations}

The model in question takes as its starting point a set of differential
equations that are, in the $N\rightarrow\infty$ limit, exact but form an infite
hierarchy. We present the first three orders of this:

\begin{eqnarray}
\frac{d}{dt} \mynode{S} & = & -\tau \mypair{S}{I} + g\mynode{I} \; , 
  \nonumber \\
\frac{d}{dt} \mynode{I} & = & \tau \mypair{S}{I} - g\mynode{I} \; , 
  \nonumber \\
\frac{d}{dt} \mypair{S}{S} & = & - 2 \tau \mythreemotif{S}{S}{I}{2}
+ 2 g \mypair{S}{I} \; , 
  \nonumber \\
\frac{d}{dt} \mypair{S}{I} & = & \tau \left( \mythreemotif{S}{S}{I}{2}
- \mythreemotif{I}{S}{I}{2} - \mypair{S}{I} \right)
+ g \left( \mypair{I}{I} - \mypair{S}{I} \right) \; , 
  \nonumber \\
\frac{d}{dt} \mypair{I}{I} & = & 2 \tau \left( \mythreemotif{I}{S}{I}{2}
+ \mypair{S}{I} \right)
- 2 g \mypair{I}{I} \; , 
  \nonumber \\
\frac{d}{dt}\mythreemotif{S}{S}{S}{0} & = &
- \, \tau \left(2\myfourmotif{S}{S}{S}{I}{2}{1}{0}{2} +
\myfourmotif{S}{S}{S}{I}{1}{2}{0}{2} \right) + g \left(
2\mythreemotif{S}{S}{I}{0} + \mythreemotif{S}{I}{S}{0} \right) \; ,
  \nonumber \\
\frac{d}{dt}\mythreemotif{S}{S}{I}{0} & = &
\tau \left(\myfourmotif{S}{S}{S}{I}{2}{1}{0}{2} -
\myfourmotif{S}{S}{I}{I}{2}{2}{0}{1} -
\myfourmotif{S}{S}{I}{I}{1}{2}{0}{2} - \mythreemotif{S}{S}{I}{0}
\right) \nonumber \\ &&  
+ g \left(  \mythreemotif{S}{I}{I}{0} + \mythreemotif{I}{S}{I}{0} 
- \mythreemotif{S}{S}{I}{0} \right) \; , 
  \nonumber \\
\frac{d}{dt}\mythreemotif{S}{I}{S}{0} & = &
+ \tau \left( \myfourmotif{S}{S}{S}{I}{1}{2}{0}{2}
-2 \myfourmotif{S}{I}{S}{I}{2}{1}{0}{2} -2\mythreemotif{S}{I}{S}{0} \right) 
\nonumber \\ 
& & + g \left( 2\mythreemotif{S}{I}{I}{0} - \mythreemotif{S}{I}{S}{0} \right)
\; , 
  \nonumber \\
\frac{d}{dt}\mythreemotif{S}{I}{I}{0} & = &
 \tau \Big(\myfourmotif{S}{I}{S}{I}{2}{1}{0}{2} +
 \myfourmotif{S}{S}{I}{I}{1}{2}{0}{2}
- \myfourmotif{S}{I}{I}{I}{2}{2}{0}{1}  
\nonumber \\ & & + \mythreemotif{S}{I}{S}{0}
+ \mythreemotif{S}{S}{I}{0} - \mythreemotif{S}{I}{I}{0} \Big)
+ g \left( \mythreemotif{I}{I}{I}{0}
- 2 \mythreemotif{S}{I}{I}{0} \right)  \; , 
  \nonumber \\
\frac{d}{dt}\mythreemotif{I}{S}{I}{0} & = &
\tau \left( 2\myfourmotif{S}{S}{I}{I}{2}{2}{0}{1} 
- \myfourmotif{I}{S}{I}{I}{1}{2}{0}{2} - 2\mythreemotif{I}{S}{I}{0} \right) 
\nonumber \\& &
+ g \left( \mythreemotif{I}{I}{I}{0} 
- 2\mythreemotif{I}{S}{I}{0} \right) \; , 
  \nonumber \\
\frac{d}{dt}\mythreemotif{I}{I}{I}{0} & = &
\tau \left(2 \myfourmotif{S}{I}{I}{I}{2}{2}{0}{1} +
\myfourmotif{I}{S}{I}{I}{1}{2}{0}{2} + 2 \mythreemotif{S}{I}{I}{0} +
2\mythreemotif{I}{S}{I}{0} \right)
\nonumber \\ & &
- \, 3g \mythreemotif{I}{I}{I}{0} \; , 
  \nonumber \\
\frac{d}{dt}\mythreemotif{S}{S}{S}{1}
& = &- \, 3\tau\myfourmotif{S}{S}{S}{I}{2}{1}{1}{2} +
3 g\mythreemotif{S}{S}{I}{1} \; ,
  \nonumber \\
\frac{d}{dt}\mythreemotif{S}{S}{I}{1}
& =  & \tau \left( \myfourmotif{S}{S}{S}{I}{2}{1}{1}{2}
-2 \myfourmotif{S}{S}{I}{I}{2}{2}{1}{1} -2 \mythreemotif{S}{S}{I}{1} 
\right) 
\nonumber \\ & & 
+ g\left( 2\mythreemotif{S}{I}{I}{1} -
\mythreemotif{S}{S}{I}{1} \right)
\; , \nonumber \\
\frac{d}{dt}\mythreemotif{S}{I}{I}{1}
& = &
\tau \left(2\myfourmotif{S}{S}{I}{I}{2}{2}{1}{1}
-\myfourmotif{S}{I}{I}{I}{2}{2}{1}{1}
+2\mythreemotif{S}{S}{I}{1} 
-2\mythreemotif{S}{I}{I}{1} \right) \nonumber \\ & &
+ g\left(\mythreemotif{I}{I}{I}{1} - 2\mythreemotif{S}{I}{I}{1} \right)
\; , 
  \nonumber \\
\frac{d}{dt}\mythreemotif{I}{I}{I}{1}
& =  &  3 \tau \left(\myfourmotif{S}{I}{I}{I}{2}{2}{1}{1} +
2\mythreemotif{S}{I}{I}{1} \right)
-  3g \mythreemotif{I}{I}{I}{1} \; .
\label{eq:sistriple}
\end{eqnarray}
Here and throughout this paper, we use dotted lines to imply expansion as
below:
\ba
\mythreemotif{A}{B}{C}{2} & = \mythreemotif{A}{B}{C}{0}
+ \mythreemotif{A}{B}{C}{1} \; , \\
\myfourmotif{A}{B}{C}{D}{2}{1}{0}{2} & =
\myfourmotif{A}{B}{C}{D}{0}{1}{0}{0} +
\myfourmotif{A}{B}{C}{D}{0}{1}{0}{1} +
\myfourmotif{A}{B}{C}{D}{1}{1}{0}{0} +
\myfourmotif{A}{B}{C}{D}{1}{1}{0}{1} \; ,\\
& \ \;\vdots
\ea
and similarly for other fourth-order terms.

\subsection{Closure schemes}

\begin{figure}[H]
\begin{center}
\scalebox{0.9}{\resizebox{\textwidth}{!}{ \includegraphics{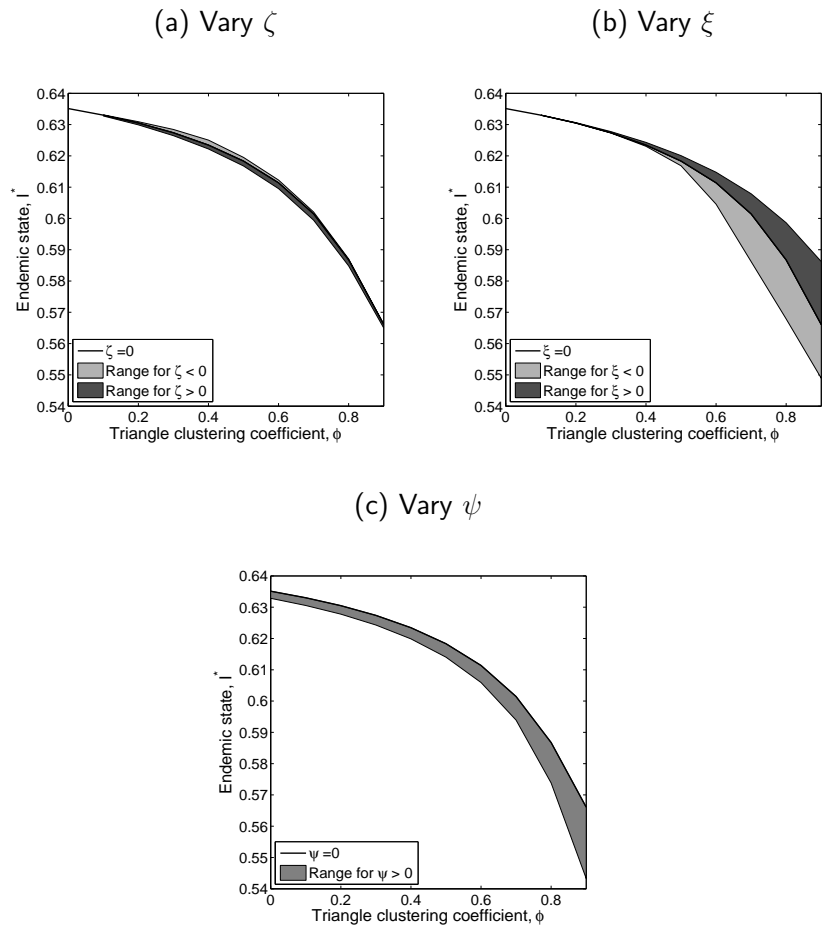} }}
\end{center}
\caption{Dynamical results for the endemic state of the triplewise contact
process model for $n=6, g=1, \tau=3/5$ and other parameters varied.
}
\label{fig:dyn}
\end{figure}

To integrate the system as presented so far, we need a closure scheme,
previously introduced in~\cite{House:2009p1790}, which is most easily expressed
in terms of the raw motif prevalences.
\begin{eqnarray}
\myfourmotif{A}{B}{C}{D}{1}{0}{0}{0} & \approx &
\num{\mtI}\frac{\num{\mtPair}^3}{\num{\mtTriple}^3 \num{\mtNode}}
\frac{\mythreemotif{A}{B}{C}{0}\mythreemotif{A}{B}{D}{0}
\mythreemotif{C}{B}{D}{0}[B]}{\mypair{A}{B}\mypair{B}{C}\mypair{B}{D}}
\; , \nonumber \\
\myfourmotif{A}{B}{C}{D}{0}{1}{0}{0} & \approx &
\num{\mtII}\frac{\num{\mtPair}}{\num{\mtTriple}^2}
\frac{\mythreemotif{B}{C}{D}{0}\mythreemotif{A}{B}{C}{0}}
{\mypair{B}{C}}
\; , \nonumber \\
\myfourmotif{A}{B}{C}{D}{0}{1}{1}{0} & \approx &
\num{\mtIII}\frac{\num{\mtPair}^3}
{\num{\mtTriangle} \num{\mtTriple}^2 \num{\mtNode}}
\frac{\mythreemotif{A}{B}{C}{1}\mythreemotif{B}{C}{D}{0}
\mythreemotif{A}{C}{D}{0}[C]}
{\mypair{A}{C}\mypair{B}{C}\mypair{C}{D}}
\; , \nonumber \\
\myfourmotif{A}{B}{C}{D}{0}{1}{0}{1} & \approx &
\num{\mtIV}\frac{\num{\mtPair}^4}{\num{\mtTriple}^4}
\frac{\mythreemotif{A}{B}{C}{0}\mythreemotif{B}{C}{D}{0}
\mythreemotif{C}{D}{A}{0}\mythreemotif{D}{A}{B}{0}}
{\mypair{B}{C}\mypair{C}{D}\mypair{D}{A}\mypair{A}{B}}
\; , \nonumber \\
\myfourmotif{A}{B}{C}{D}{1}{1}{1}{0} & \approx &
\num{\mtV}\frac{\num{\mtPair}}{\num{\mtTriangle}^2}
\frac{\mythreemotif{B}{C}{D}{1}\mythreemotif{A}{B}{C}{1}}
{\mypair{B}{C}}
\; , \nonumber \\
\myfourmotif{A}{B}{C}{D}{1}{1}{1}{1} & \approx &
\num{\mtVI}\frac{\num{\mtPair}^6}{\num{\mtTriangle}^4\num{\mtNode}^4}
\times \nonumber \\ & &
\frac{\mythreemotif{A}{B}{C}{1}\mythreemotif{B}{C}{D}{1}
\mythreemotif{C}{D}{A}{1}\mythreemotif{D}{A}{B}{1}[A][B][C][D]}
{\mypair{B}{C}\mypair{C}{D}\mypair{D}{A}\mypair{A}{B}
\mypair{A}{C}\mypair{B}{D}} \; .
\end{eqnarray}
Then~\eqref{newparams}, together with this closure and
equations~\eqref{eq:sistriple} create an integrable ODE system. Provided $\tau$
is sufficiently large compared to $g$, this system has a steady state with a
non-trivial proportion $I^*$ of the network infectious. Standardly, this
equilibrium value is called the \textit{endemic state}. We investigate this
dynamically in Figure~\ref{fig:dyn}, where $\phi$ is increased for other
parameters held constant in all plots, giving the common black line. We then
modify either (a) $\zeta$, (b) $\xi$ or (c) $\psi$. This shows that,
essentially, $\psi$ has a significant but relatively stable effect in reducing
the endemic state at each $\phi$ value, while $\xi$ can have a significant
effect in either direction at larger $\phi$ values. $\zeta$, on the other hand,
is relatively dynamically unimportant, except perhaps at moderate values of
$\phi$.  We also note that positive values of $\zeta$ reduce the endemic state,
and negative values increase it, while the opposite is true for $\xi$.

\section{Discussion}

This paper has presented a novel way of thinking about higher-order structure
in networks, together with intuitive explanations of this, rewiring schemes and
dynamical consequences. This opens up three main questions.

Firstly, what are reasonable parameter values for networks that are seen in
nature, and which can be explicitly constructed? The exact values of clustering
coefficients considered in the \SIS{} model are perhaps slightly larger than
are likely to be seen or constructed, although this should be mainly of
quantitative importance since the qualitative dynamical implications found for
higher order clustering are not modified at different coefficient values, and
moment closure (particularly in the three-motif case) has been extensively used
in modelling \SIS{} and \SIR{} dynamics without producing qualitatively
incorrect results~\cite{KRbook}. Nevertheless, if sufficiently efficient
methods were available to generate explicit networks to run stochastic
simulations on, that would significantly increase the confidence in the results
obtained here using moment closure.

Secondly, how can this analysis be generalised to networks with heterogeneous
numbers of links, and (perhaps more problematically) preferential assortative
connection between nodes of similar degree? While such analysis is doubtless
possible, the large number of interacting quantities may make analytic results
technically difficult. In particular, it it unlikely that arbitrary
heterogeneity and clustering statistics are compatible with each other.

Finally, under what conditions is it necessary to consider $k$-motifs for a
given $k$? Clearly, a high preponderance of triangles in a network would favour
a pairwise model, but this answer is less clearly posed for four-motifs in
general. However, the parameterisation suggested here goes some way towards
answering this question: starting with a set of four-motif prevalences, are
these significantly different from what would be predicted based on the values
of three- two- and one-motif parameters $\phi$, $n$ and $N$? If the new
generalised clustering parameters $\phi$, $\zeta$ and $\xi$ are significantly
different from zero, then we would expect that at least four-motifs should be
considered in analysis of the network.

\section*{Acknowledgments}

Work funded by the UK Medical Research Council (Grant Number G0701256) and the
UK Engineering and Physical Sciences Research Council (Grant Number
EP/H016139/1).  The author would like to thank Matt Keeling and Matthew Vernon
for helpful discussions during the preparation of this manuscript.

\newpage


\end{document}